\documentclass[11pt,preprint]{aastex}

\pdfoutput=1
\shorttitle{Calibration of $BVRI$ Photometry for ACS/WFC}
\shortauthors{Saha et al.}

\begin{document}

\title{Calibration of $BVRI$ Photometry for the Wide Field Channel of the \textit{HST} Advanced Camera for Surveys }

\author{Abhijit Saha\altaffilmark{1}, Richard A.\ Shaw\altaffilmark{1},  Jennifer A.\ Claver\altaffilmark{1}, and Andrew~E.\ Dolphin\altaffilmark{2}}
\email{saha@noao.edu, shaw@noao.edu, jenna@noao.edu, adolphin@raytheon.com}

\altaffiltext{1}{National Optical Astronomy Observatory, Tucson, AZ  85719, USA. NOAO is operated by the Association of Universities for Research in Astronomy, Inc. (AURA) under cooperative agreement with the National Science Foundation}
\altaffiltext{2}{Raytheon Company, 1151 E.\ Hermans Rd., Tucson, AZ  85756, USA.}

\begin{abstract}
We present new observations of two Galactic globular clusters, PAL4 and PAL14, using the Wide-Field Channel of the Advanced Camera for Surveys (ACS) on board the \textit{Hubble Space Telescope} (\textit{HST}), and reanalyze archival data from a third, NGC2419. We matched our photometry of hundreds of stars in these fields from the ACS images to existing, ground-based photometry of faint sequences which were calibrated on the standard $BVRI$ system of Landolt. These stars are significantly fainter than those generally used for \textit{HST} calibration purposes, and therefore are much better matched to supporting precision photometry of ACS science targets. We were able to derive more accurate photometric transformation coefficients for the commonly used ACS broad-band filters compared to those published by \citet{Sir05}, owing to the use of a factor of several more calibration stars which span a greater range of color. We find that the inferred transformations from each cluster individually do not vary significantly from the average, except for a small offset of the photometric zeropoint in the F850LP filter. Our results suggest that the published prescriptions for the time-dependent correction of CCD charge-transfer efficiency appear to work very well over the $\sim3.5$~yr interval that spans our observations of PAL4 and PAL14 and the archived images of NGC2419. 
\end{abstract}


\section{Introduction}

Since its installation on the \textit{Hubble Space Telescope} (\textit{HST}) in 2002 March, the Advanced Camera for Surveys (ACS) \citep{Ford_etal03} has been used extensively for a wide variety of science programs. Many programs place high demands on the accuracy of the photometric calibration, and on the accuracy of the transformation of the native ACS system to standard systems. In this study we present observations and analysis that lead to a reliable mapping from magnitudes measured on the native system of the ACS Wide Field Channel (WFC) to magnitudes in the $BVRI$ passbands of the \citet{Landolt83, Landolt92} system. 
Two considerations have fuelled this study. First, much of the legacy in stellar broad-band photometry in the visible spectral region is published in the Landolt system, which is itself a continuation and consolidation of the earlier Johnson-Kron-Cousins system. For instance, the Cepheid period-luminosity relations are almost always quoted in terms of Landolt magnitudes, and to derive accurate corrections for extinction one must work in colors defined from Landolt magnitudes. The ability to map ACS observations to the Landolt system accurately is thus central to distance scale work with ACS. Work on stellar populations in the Milky Way and external galaxies, including ages and metallicities of clusters and integrated stellar fields, are also fundamentally based on Landolt photometry. 
Second, most \textit{HST} programs observe objects that are very faint, yet the typical approach to calibration, largely in the interest of calibrating many operating modes in a short time, is to obtain short exposures of brighter standard stars. While the detectors (CCDs, in the case of the WFC) have very linear response over a large dynamic range, problems such as imperfect charge transfer efficiency (CTE) can seriously degrade the accuracy of the calibration for faint targets in long exposures. For precision photometry, calibrators and science targets should be observed in the same way to the extent possible. 
The calibration study presented here uses three well studied, distant Galactic globular clusters: \object{NGC2419}, \object{PAL4} and \object{PAL14}, each with star sequences calibrated to $V \sim 22$~mag. The comparison against these stars, which are within $\sim 5$ magnitudes of most ACS science targets, is much more robust against uncertainties in non-linearity and CTE corrections than with standard stars that are brighter by 5~mag or more. 

The on-orbit photometric performance and calibration of the ACS (in both the WFC as well as the High Resolution 
Channel\footnote{The HRC, which was never as heavily used as the WFC, failed in 2007 January.}) 
was presented in \citet[][hereafter S05]{Sir05}, 
which remains the key reference for this instrument. These photometric calibrations covered multiple photometric systems native to \textit{HST} instruments, as well as transformations to the Landolt system. The S05 transformations were derived from a combination of synthetic photometry, based upon pre-launch tabulated responses of the individual component optics, as well as direct empirical comparisons with stars in the very metal poor globular cluster NGC2419. The aim of the synthetic photometry was to cover a wider range of source spectral energy distributions and foreground interstellar reddening than is normally sampled in the ACS calibration fields. 
Our study focusses on the WFC and presents the comparison with two additional targets, PAL4 and PAL14, each with their own sequences. Our study is entirely empirical, and includes a much larger sample of calibration stars that span at least 5~mag in brightness. 
The ensemble of stars also spans a range of source colors wide enough to 
significantly reduce the uncertainties in the color term of the derived transformations. 
We inter-compare the results for each target to provide an important external measure of the errors. 
In addition, the two new observation sets were obtained 3.5 years or more after the NGC2419 observations, a time baseline sufficient to reveal temporal changes in instrument performance. We are thus able to evaluate the accuracy of the prescriptions for CTE correction provided by the ACS instrument team at STScI \citep{chia09}, which are extremely important for this instrument because of the large number of parallel and serial shifts that are required to read out the WFC detectors \citep{acs_ihb10}.
We present our new observations in \S2, and photometric measurements on the calibrator stars in \S3. In \S4 we describe our derivation of the photometric  transformations themselves, and we present the results in tabular and graphical form. We conclude in \S5 with a discussion of our error analysis, the applicability of our transformations, and of the realized improvement in the ACS/WFC photometric calibration. 

\section{Observations}

We observed the Galactic globular clusters PAL4 and PAL14 (GO program 10622) using the ACS/WFC in ACCUM mode and the default CCD gain of 2. We obtained paired (CR-SPLIT) images in each of the nine medium- and broad-band filters that span the full spectral range of the WFC, with total exposure times sufficient to yield a signal-to-noise ratio (S/N) of $\gtrsim10$ for the faintest photometric standards in these fields. 
We also analyzed archival images of the globular cluster NGC2419, which were obtained as part of the early, on-orbit ACS calibration program 9666 (see S05). These images were also obtained in ACCUM mode with a gain of 2; we restricted our analysis to the WFC images and the 8 broad-band filters in common with our program. Details of the observations may be found in Table~\ref{obslog}.

We performed our analysis on calibrated ACS images as furnished by the standard ACS processing pipeline (CALACS version 5.0.5, 2009 Aug.); specifically, images with the suffix ``\_drz'' in the filename. While the use of standard products from the HST archive yields calibration products with the widest applicability, we did process the raw images with our own methods as a check: as expected, photometry of images produced in this way does not show significant differences from that derived from the calibrated images served by the archive. The processing steps in CALACS were described by S05, and are documented in detail by \citet{acs_dhb06}. In brief, the instrumental signature is removed from the raw data by subtracting the bias level and residual bias structure, removing the overscan regions, scaling and subtracting a master dark frame, normalizing to unit gain, and correcting for photometric uniformity (including pixel-to-pixel sensitivity variations within a CCD, detector-to-detector response normalization, and the focal plane illumination pattern) by dividing out a master flat-field. Cosmic rays are rejected when the CR-split images for each filter are co-added. The combined images are geometrically rectified and the brightness is normalized unit exposure time. 

\section{Measuring Instrumental Magnitudes}
\label{measmag}

The analysis of the combined, calibrated images begins with object identification and the determination of instrumental magnitudes for all stars in each passband. We used a modified version of the \textit{DoPHOT} program \citep{Sch93} to perform point-spread-function (PSF) fitted stellar photometry, where the PSF is held constant with position in the field of view. Our version of \textit{DoPHOT} also produces \textit{aperture magnitudes} over a range of radii for the brighter and relatively uncrowded stars {\it in isolation} (i.e. with all other objects subtracted after fitting); the approach is similar to that described by \citet[][hereafter SDTW]{Saha05}. From these brighter, isolated stars we first determine the field dependence (and constant offset, if any) from our measured PSF magnitudes to small-aperture magnitudes. A final correction, from small-aperture measurements to infinite-aperture magnitudes, can be derived from a knowledge of the PSF encircled energy curves, as we describe below. 

Let  $m_{fit}$ denote the reported PSF fitted magnitude, and $m_{ap}$ the magnitude as obtained through a finite aperture. The aperture correction is then given by $apcorr = m_{ap} - m_{fit}$, and is computed for the isolated, high S/N objects for which $m_{ap}$ are available. The resulting individual values of $apcorr$ (each with associated error estimates propagated from reported error estimates for the individual $m_{fit}$ and $m_{ap}$) were examined to search for dependence with position of the star in the field of view (FoV).  Such variation can arise because of subtle variations in the PSF across the FoV, due to the changing alignment of the camera and telescope optics across focal plane. With experimentation using a variety of instrumentation, we have found that a quadratic polynomial with circular symmetry about the center of the FoV is an adequate representation\footnote{A general two dimensional quadratic function has the risk of being unconstrained in the corners if there are not enough stars to define it.}, 
provided an additional term is included to correct the discontinuity going from one detector to the next. In the case of the ACS/WFC, the effects were almost unnoticeable.  Once the surface polynomial description of $apcorr$  (in terms of position on the FoV) is evaluated as described above, all of the PSF fitted magnitudes for stars anywhere in the FoV can be put on the system of $m_{ap}$. Software necessary to determine and apply the aperture corrections as above was custom written in the {\it IDL} language.
The advantage of the above aperture measurement scheme is that we measure the star in isolation (with all neighbors subtracted), and do so automatically for all the stars (from several to several hundreds, depending on the target field), that have adequate S/N. The disadvantage is that an automatic algorithm for flat growth curve background determination can go awry from specious or unsubtracted features in the image.  However, this problem is mitigated by the statistics from a large number of stars, and also the reported errors in the aperture measurements account for misbehavior of the growth curves. 

What remains is to apply the correction from small-aperture to infinite-aperture magnitudes. Specifically, consider the case where the background is set to that value for which the growth curve is flat for apertures of size ranging from 12 to 16~pixel radius (0\farcs6 to 0\farcs75 on the rectified WFC images), and designate the energy above this background (which includes the wings of the stellar PSF) within an aperture of 10~pixel (0\farcs5) radius by $E^{\prime}(10)$. We can obtain the total energy from the source  from $E^{\prime}(10)$, given prior knowledge of the PSF encircled energy as a function of radius.
If $e(n)$  denotes the  true encircled energy from a point source within an aperture of radius $n$ pixels, normalized to unit total energy from the source,  then it can be shown that 
\begin{equation}
E^{\prime}(10) =  E(\infty) \times \left[ e(10) - \frac{e(16) - e(12)}{f} \right]
\end{equation}
where $E(\infty)$ is the total intensity of the source. For each filter of interest the values of $e(10)$, $e(12)$ and $e(16)$ are given in Table~3 of S05, and the scalar $f$ within the square brackets in the above equation is a multiplicative constant of order unity that changes only from one pass-band to another.  We denote the magnitude as corrected for $E(\infty)$ by $m_{inst}$. 

Charge transfer losses in the instrument detectors are a function of the object flux as well as the background (sky) exposure levels. In addition, the effects are known to worsen over time. A study of these effects are reported by \citet{chia09}, where they demonstrate that the effect on photometry can be well characterized.  Their Equation (2) and acompanying Table~2 provide a prescription for correcting the extracted photometry, which is a function of the measured accumulated flux in the object, the exposure level of the background, and the epoch of observation. Our study adopts their quantitative characterization: the $m_{inst}$ values described above were corrected according to the \citet{chia09} scheme. Since our data for PAL4 and PAL14 were obtained  more than three years after the NGC2419 data, the comparison of the calibration zero-points in the data from the separate epochs must show consistency if the adopted CTE correction scheme is correct. The magnitude of the CTE correction to the full sample is significant: for the extreme case of a single star with a detected flux of 100~$e^-$ above a negligible sky that suffers the largest possible number of parallel shifts on the CCD, we would expect a correction of $\sim0.3$~mag over the interval between epochs of the observations. However, since all stars used in the derivation of these zero-points in this paper are based on relatively bright stars, this is a necessary, but not sufficient criterion for validating the scheme. 

\begin{figure}
\epsscale{.60}
\plotone{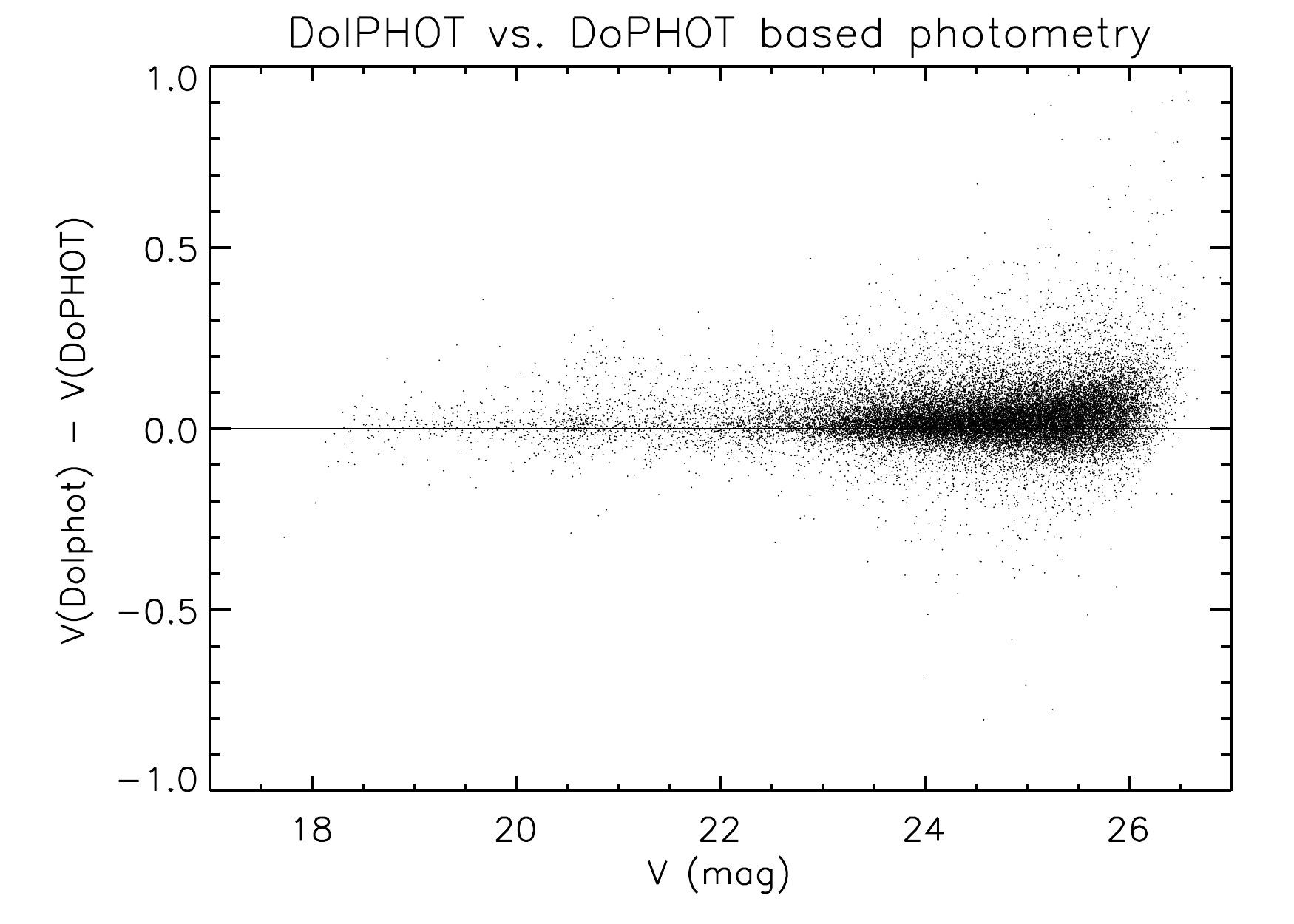}
\caption{Comparison of photometry as produced by our version of DoPHOT \citep{Sch93} and DOLPHOT \citep{Dolphin00} for the $V$ passband. 
\label{fig:PhotCmp}}
\end{figure}

As a check on the photometric technique described above, we reduced the photometric data with independent software, and performed photometry on those images using 
DOLPHOT\footnote{DOLPHOT is based on HSTphot, which is available at: \url{http://purcell.as.arizona.edu/dolphot/}.}, 
the algorithms for which are based upon HSTphot and are described in detail by \citet{Dolphin00}. This program employs PSF-fits to stars, and employs PSFs as computed over a spatial grid for ACS with 
TinyTim\footnote{TinyTim PSF modelling software is available at \url{http://www.stsci.edu/software/tinytim/tinytim.html}.} 
software. DOLPHOT also works directly on calibrated, CR-SPLIT images that have \textit{not} been combined and geometrically rectified with the Drizzle software used in the CALACS pipeline (i.e., images with the filename suffix ``\_flt"). The measured magnitudes from both programs were corrected for CTE loss. 
We show a comparison in Figure~\ref{fig:PhotCmp}, where the overall the agreement is excellent, particularly for stars with $V<23$. The scatter in the relation is generally well within $\sim0.1$ mag in this range, as expected. Fainter than this the DOLPHOT magnitudes begin to exceed those of DoPHOT, with the trend growing to $\sim0.03$~mag at $V=26$. We expect the scatter in the relation to be inflated at the faint end relative to statistical deviations for two reasons. First, the selection of type-1 targets (stars) in DOLPHOT from all available targets (stars + galaxies) is effectively a faint limit cut, which would make the DoPHOT magnitudes trend a little brighter. 
Second, using resampled images in DoPHOT introduces a systematic error due to correlated noise (in addition to random errors due to randomly-phased PSF mismatches), of an order comparable to the systematics that arise from the formulation of the noise term in the PSF-fitting. Taken together, the discrepancy between magnitudes determined with these techniques for data in common is quite small, and increases the confidence in the results presented in the next Section. 

\section{Transformation to the Landolt $BVRI$ System} 

The primary objective of this study is to derive the transformations from the ACS/WFC instrumental photometric system to the \citet{Landolt92, Landolt83} $BVRI$ system. For this we make use of the high quality $BVRI$ photometry for these target globular clusters from SDTW, who tabulated magnitudes for those stars for which the {\it random} errors in measurement are less than 0.015~mag in all four bands. 
We generated catalogs of photometric measurements and star positions from the ACS/WFC images of each target cluster, and matched their astrometric positions to the SDTW catalogs using an initial spatial tolerance of 1\farcs0. We refined the spatial matches by selecting the closest match when multiple candidates fell within the spatial tolerance, providing that such matches were consistent with the global trend of ACS/WFC-to-target magnitude. 

By construction, the system of $m_{inst}$ described in \S3 above corresponds to ``OBMAG" as defined in Equation~(1) of S05, and also to ``SMAG" in Equation~(12) of the same paper. We will use the term $SMAG$ to denote native instrumental magnitude in the remainder of this paper. Following the S05 notation, consider the magnitude $TMAG$, and color $TCOL$ for a star in any desired (i.e., target) photometric system. We wish to derive the correspondence: 
\begin{equation}
 TMAG = SMAG + c_0 + c_1 \times TCOL
 \label{transdef}
 \end{equation}  
Higher order terms in $TCOL$ can be added if necessary, but are not found to be needed for the data in this study. 
We performed a least-squares fit over all cataloged stars to evaluate $c_0$ and $c_1$ for each ACS filter, selecting $TMAG$s with the greatest overlap in bandpass. Relations were generated for up to six different color combinations in $TCOL$: $B-V$, $B-R$, $B-I$, $V-R$, $V-I$, and $R-I$. The fits were weighted by the inverse variance of the combined photometric uncertainties for each star. 
The results are given in Table~\ref{tab:PhotTrans} and illustrated in Figures~\ref{fig:f435w} through~\ref{fig:f850lp}. The table and the figures describe the most appropriate transformations for each ACS/WFC passband. The first three columns give the ACS/WFC filter for each transformation, followed by the target passband in the Landolt system, and the target color. The fourth column identifies the transformation coefficients, the values for which are listed in subsequent columns. The coefficients are first given separately for each observed cluster in succeeding columns, followed by the summary transformation from combining all three clusters under the heading ``Combined." The last column gives the corresponding coefficients from Table~22 of S05\footnote{The $c_2$ coefficients of S05 were always zero for transformations derived from observed (rather than synthetic) data, which is the point of comparison here.}, where available.
The first row of the table shows for the F435W filter that 211, 42, and 58 matched objects were used to derive the transformations for NGC2419, PAL4, and PAL14, respectively. Next there is a block of four rows, showing the mean color for the sample, followed by the regression (in this case) for $B-V$ as the color $TCOL$, i.e. for the equation: 
\begin{equation}
B  = SMAG(F435W) + c_0 + c_1\times(B-V)
\end{equation}
The \textit{mean color} is the weighted mean of the stars used in the fit. The values of  $c_0$ and $c_1$ for the specific choice of color, $(B-V)$ in this case, and the uncertainties associated with them are  shown in the next two rows. The coefficients were determined through linear regressions with weighting, using the inverse variance of the individual reported errors. The fourth row of the block shows what the zero-point would be were it calculated not at zero color, but for a color near the mean color. This value is insensitive to errors made in the estimation of the color term $c_1$, and is better for comparing the zero-points across all three objects and with the S05 zero-points from NGC2419. 
We were unable to re-construct the correct uncertainties for this row for the S05 values because we do not know the correlation between their quoted uncertainties for $c_0$ and for $c_1$.  We expect that they are of the order of their uncertainties in $c_0$ and $c_1$, which are typically a little larger than the corresponding uncertainties for values derived in this paper. 

We caution the reader that when applying the coefficients in Table~\ref{tab:PhotTrans} using Equation (2), the $SMAG$ term is the magnitude as measured in the source (ACS/WFC) system, while $TMAG$ and $TCOL$ are in the target (Landolt) system. Therefore $TCOL$ must be derived iteratively from observations in more than one ACS/WFC broad-band filter unless the color of the source in the Landolt system is known. 

\section{Discussion}
\subsection{Internal and External Consistency}

\begin{figure}
\epsscale{.80}
\plotone{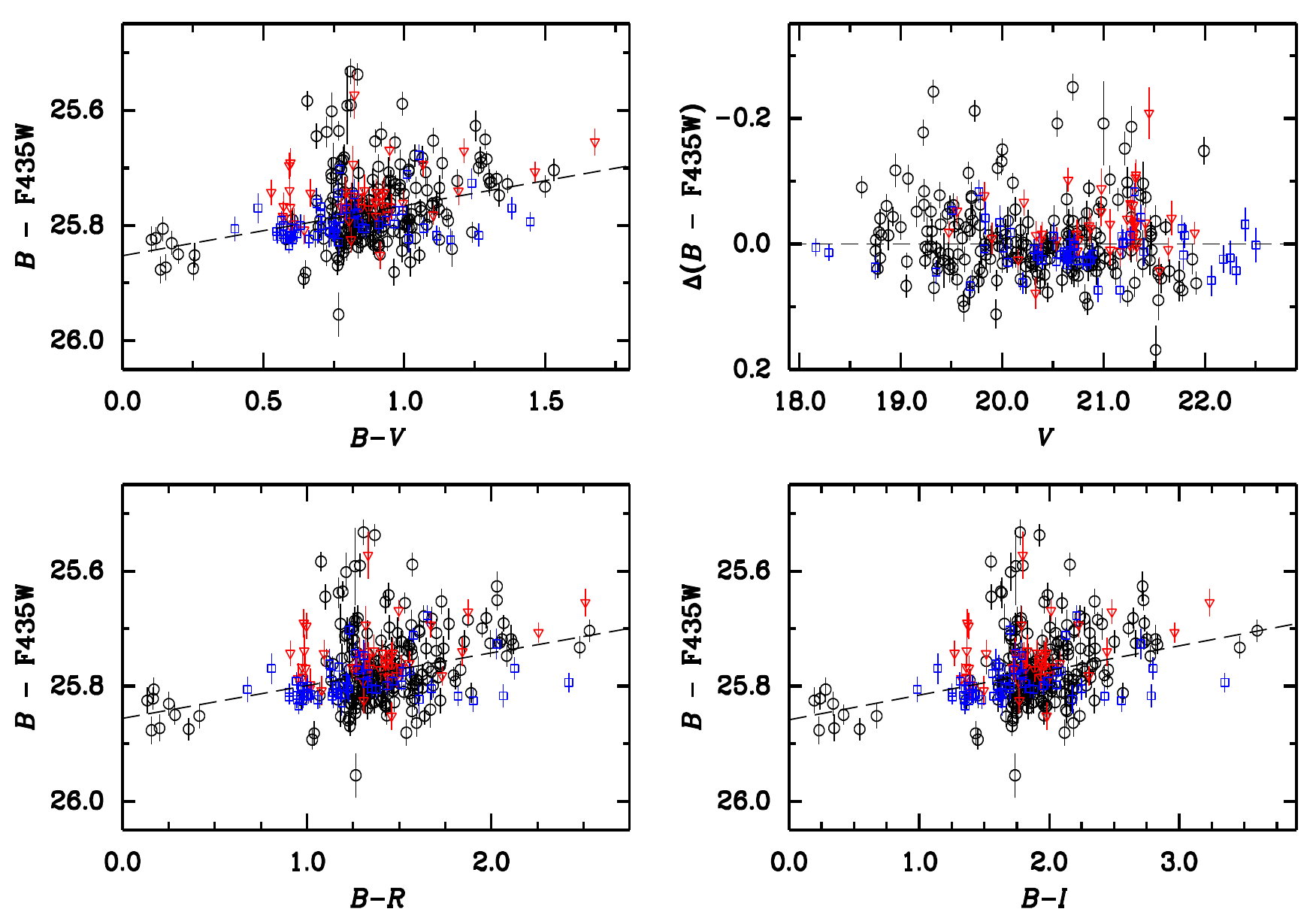}
\caption{Transformations between magnitudes in the system of \citet{Landolt83, Landolt92} and those measured with the ACS/WFC instrument and the F435W filter. Symbols denote stars from different clusters: NGC2419 (\textit{black open circles}), PAL~4 (\textit{red open triangles}), and PAL~14 (\textit{blue open squares}). Regression lines from the coefficients in Table~\ref{tab:PhotTrans} are indicated (\textit{dashed lines}). Also shown are the residuals (\textit{upper right}) from the $B-F435W$ vs. $B-V$ transformation. 
See the electronic edition of the \textit{PASP} for a color version of this figure.
\label{fig:f435w}}
\end{figure}

\begin{figure}
\epsscale{.80}
\plotone{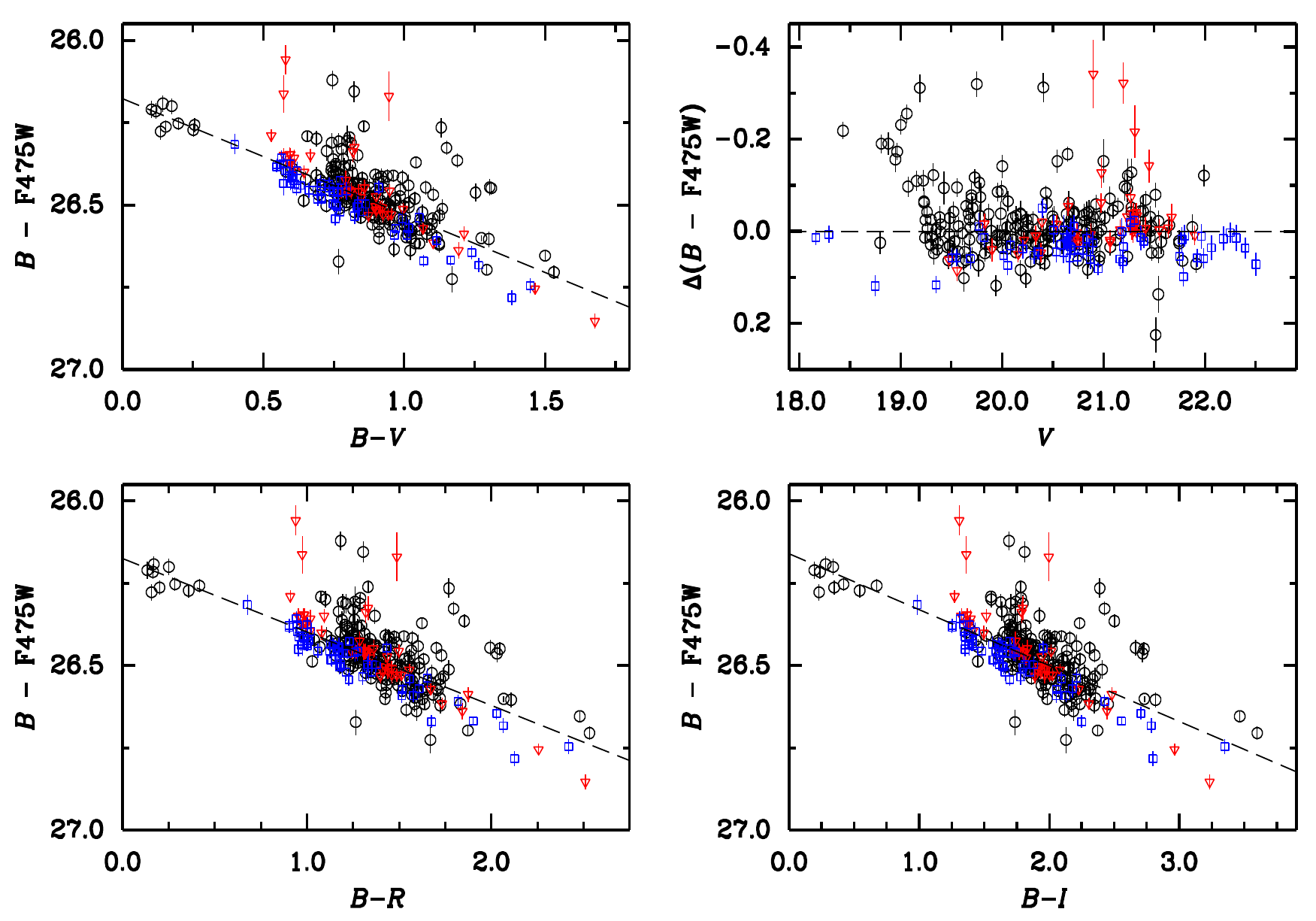}
\caption{Same as Figure~\ref{fig:f435w} for the filter F475W. Residuals refer to the $B-F475W$ vs. $B-V$ transformation. 
See the electronic edition of the \textit{PASP} for a color version of this figure.
\label{fig:f475w}}
\end{figure}

\begin{figure}
\epsscale{.80}
\plotone{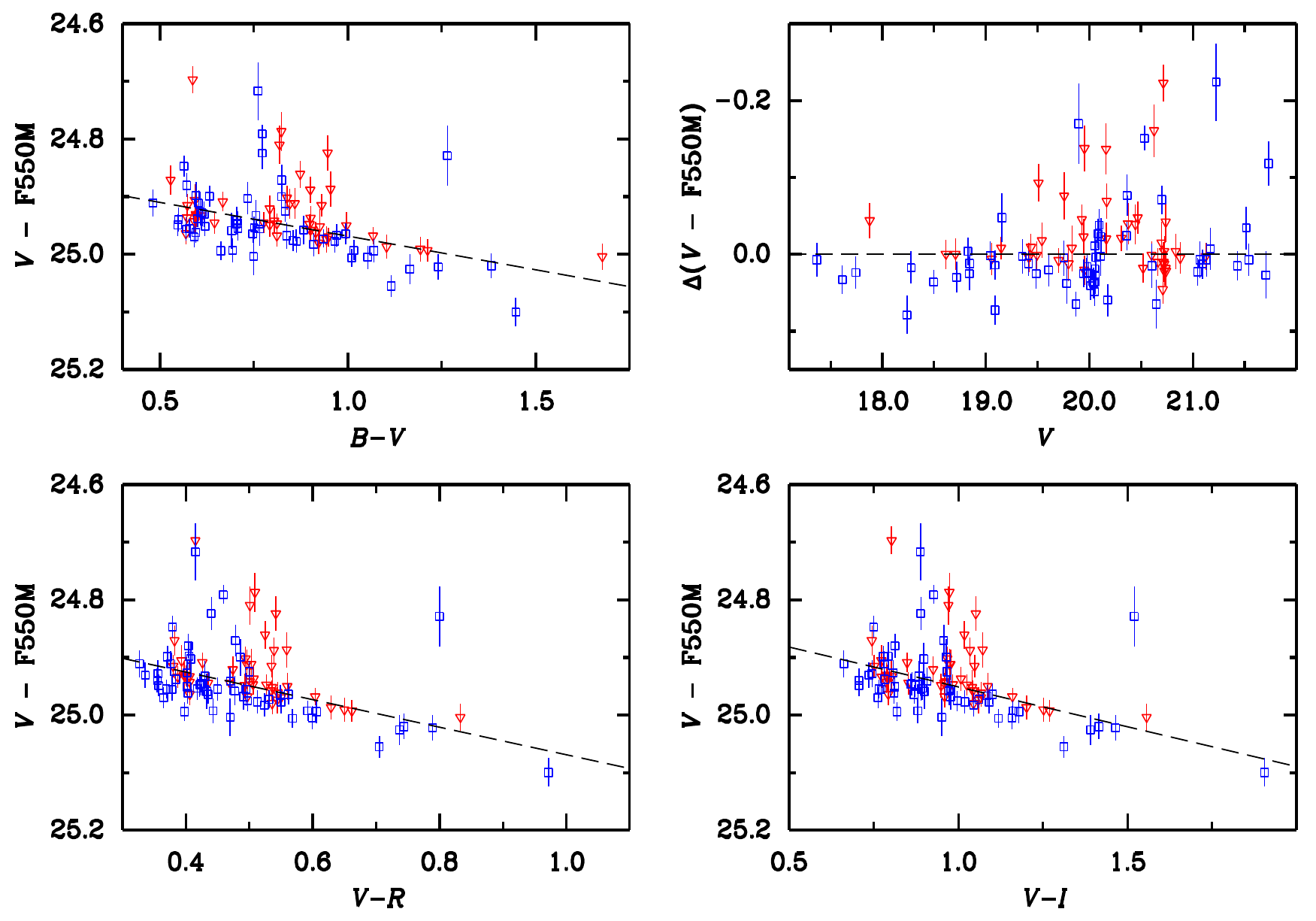}
\caption{Same as Figure~\ref{fig:f435w} for the filter F550M. Residuals refer to the $V-F550M $ vs. $B-V$ transformation.
See the electronic edition of the \textit{PASP} for a color version of this figure.
\label{fig:f550m}}
\end{figure}

\begin{figure}
\epsscale{.80}
\plotone{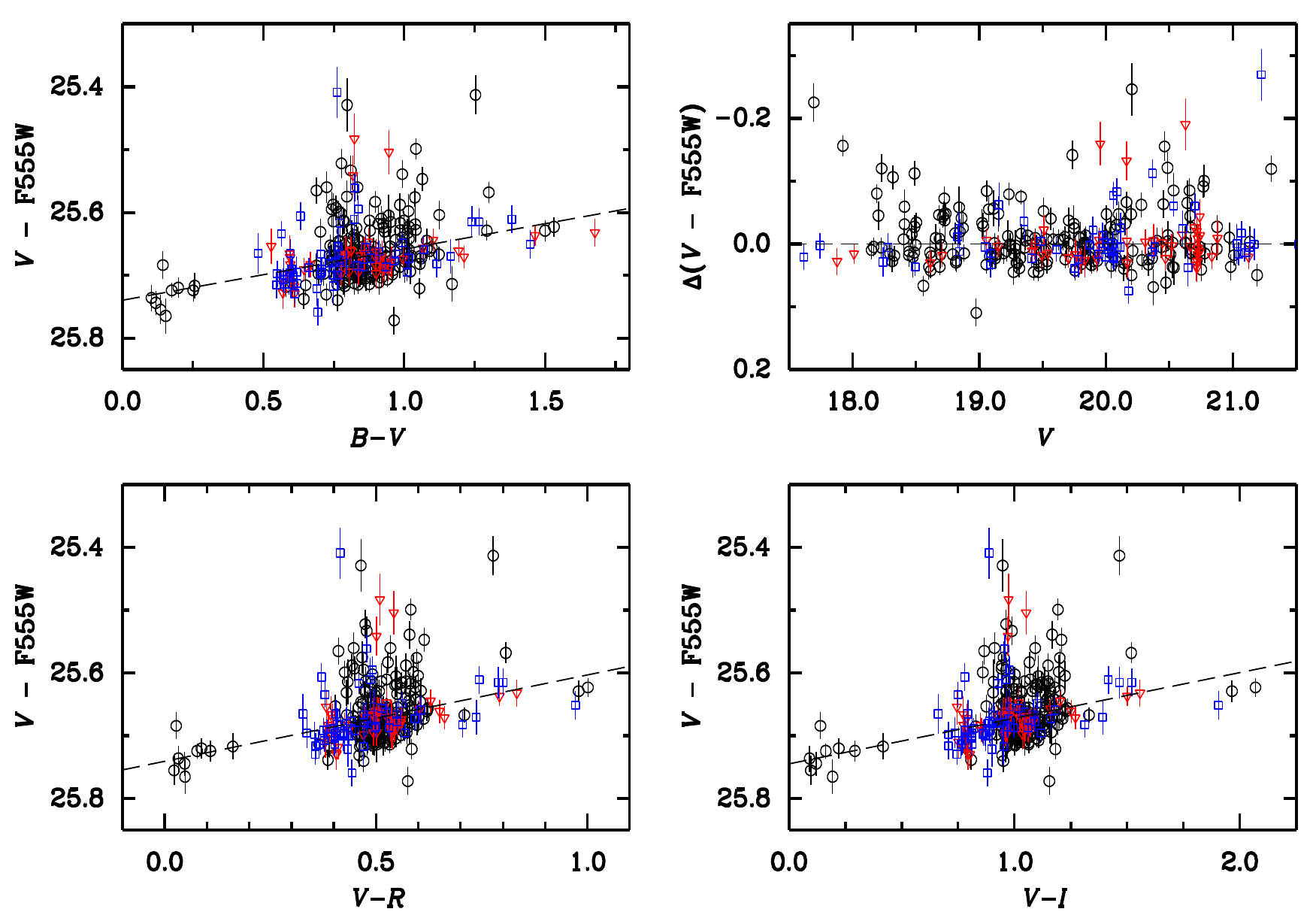}
\caption{Same as Figure~\ref{fig:f435w} for the filter F555W. Residuals refer to the $V-F555W $ vs. $B-V$ transformation. 
See the electronic edition of the \textit{PASP} for a color version of this figure.
\label{fig:f555w}}
\end{figure}

\begin{figure}
\epsscale{.80}
\plotone{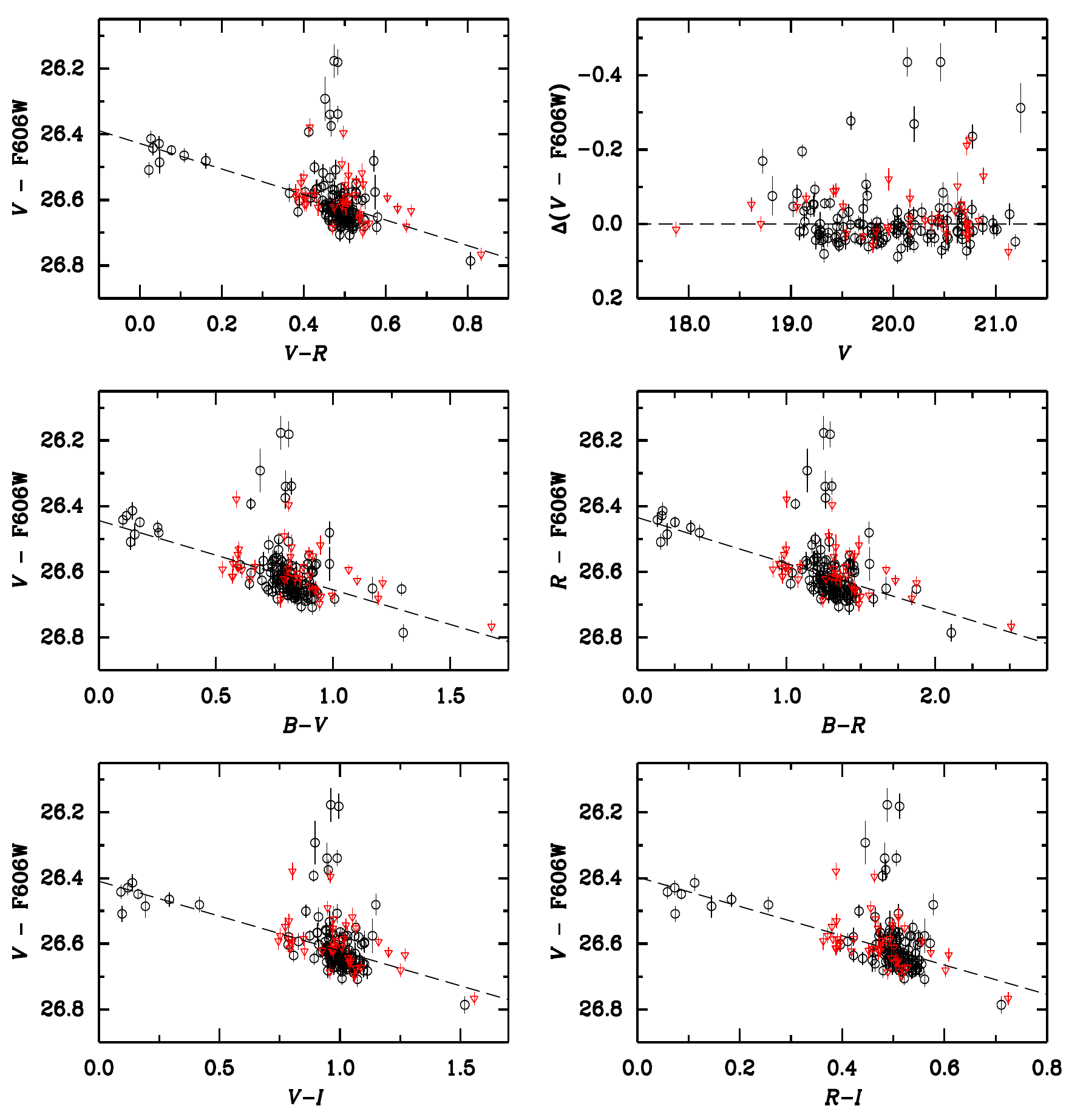}
\caption{Same as Figure~\ref{fig:f435w} for the filter F606W. Residuals refer to the $V-F606W$ vs. $V-R$ transformation. 
See the electronic edition of the \textit{PASP} for a color version of this figure.
\label{fig:f606w}}
\end{figure}

\begin{figure}
\epsscale{.80}
\plotone{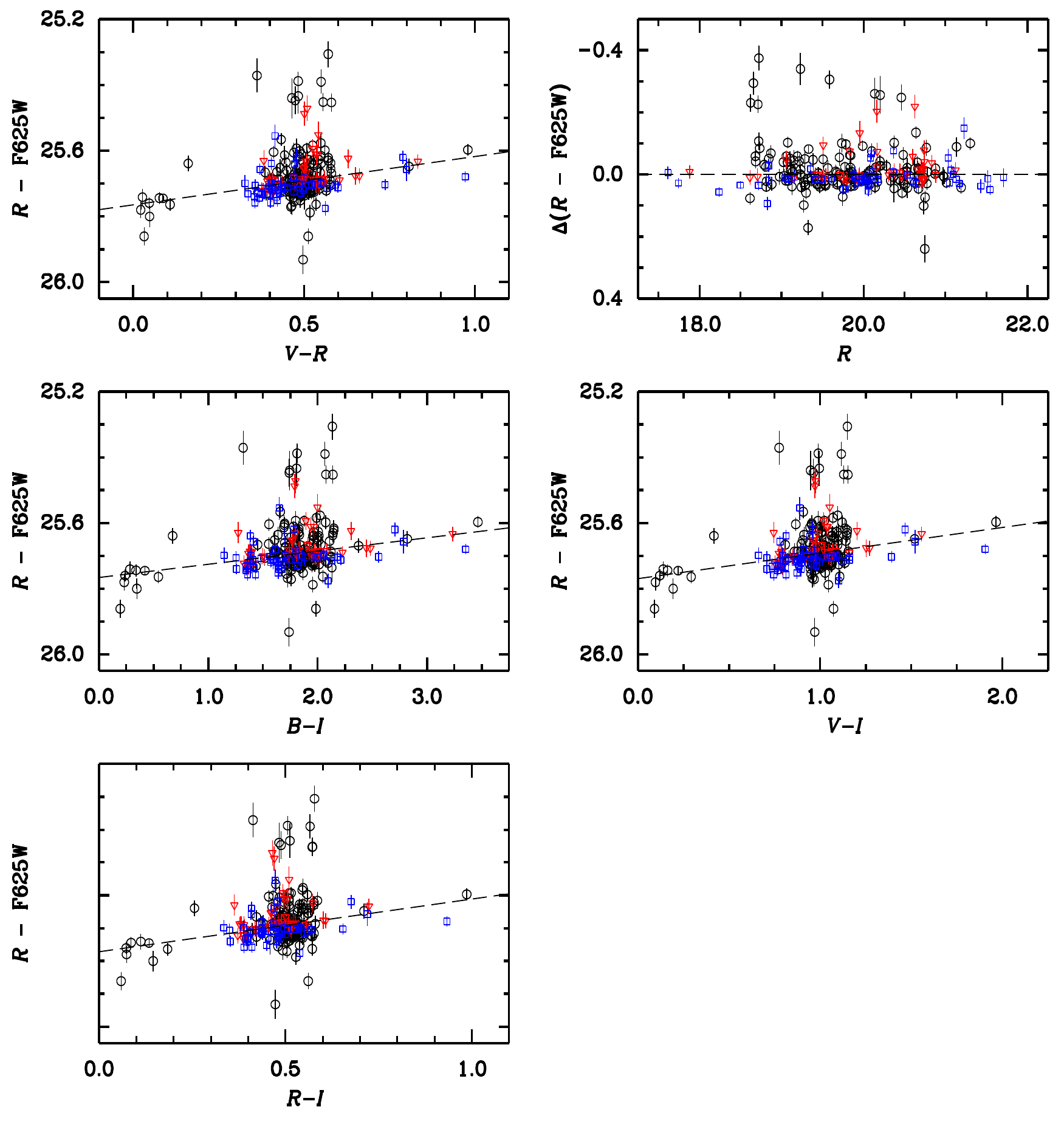}
\caption{Same as Figure~\ref{fig:f435w} for the filter F625W. Residuals refer to the $V-F625W$ vs. $V-R$ transformation.
See the electronic edition of the \textit{PASP} for a color version of this figure.
\label{fig:f625w}}
\end{figure}

\begin{figure}
\epsscale{.80}
\plotone{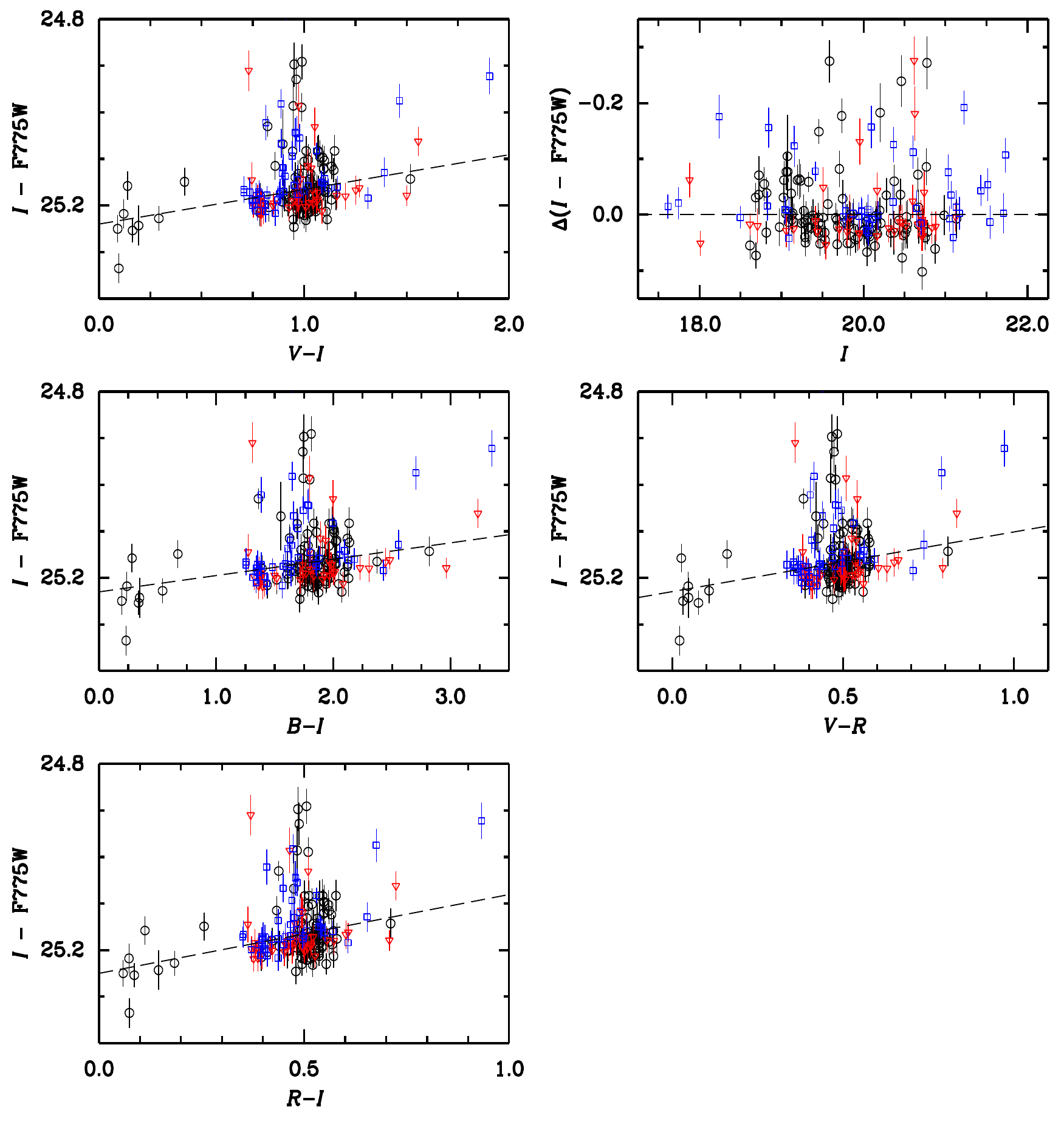}
\caption{Same as Figure~\ref{fig:f435w} for the filter F775W. Residuals refer to the $I-F775W$ vs. $V-I$ transformation.
See the electronic edition of the \textit{PASP} for a color version of this figure.
\label{fig:f775w}}
\end{figure}

\begin{figure}
\epsscale{.80}
\plotone{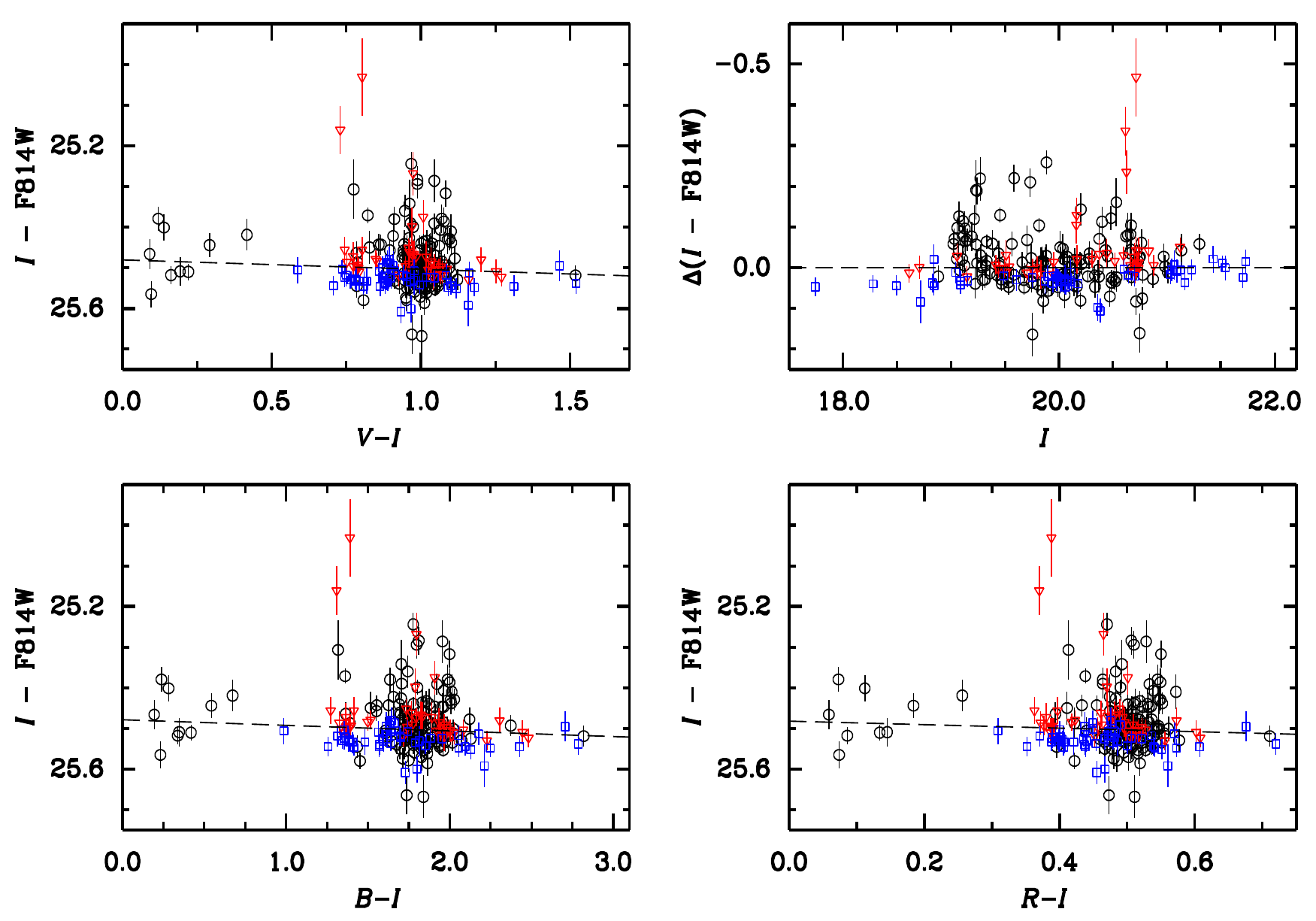}
\caption{Same as Figure~\ref{fig:f435w} for the filter F814W. Residuals refer to the $I-F814W$ vs. $V-I$ transformation.
See the electronic edition of the \textit{PASP} for a color version of this figure.
\label{fig:f814w}}
\end{figure}

\begin{figure}
\epsscale{.80}
\plotone{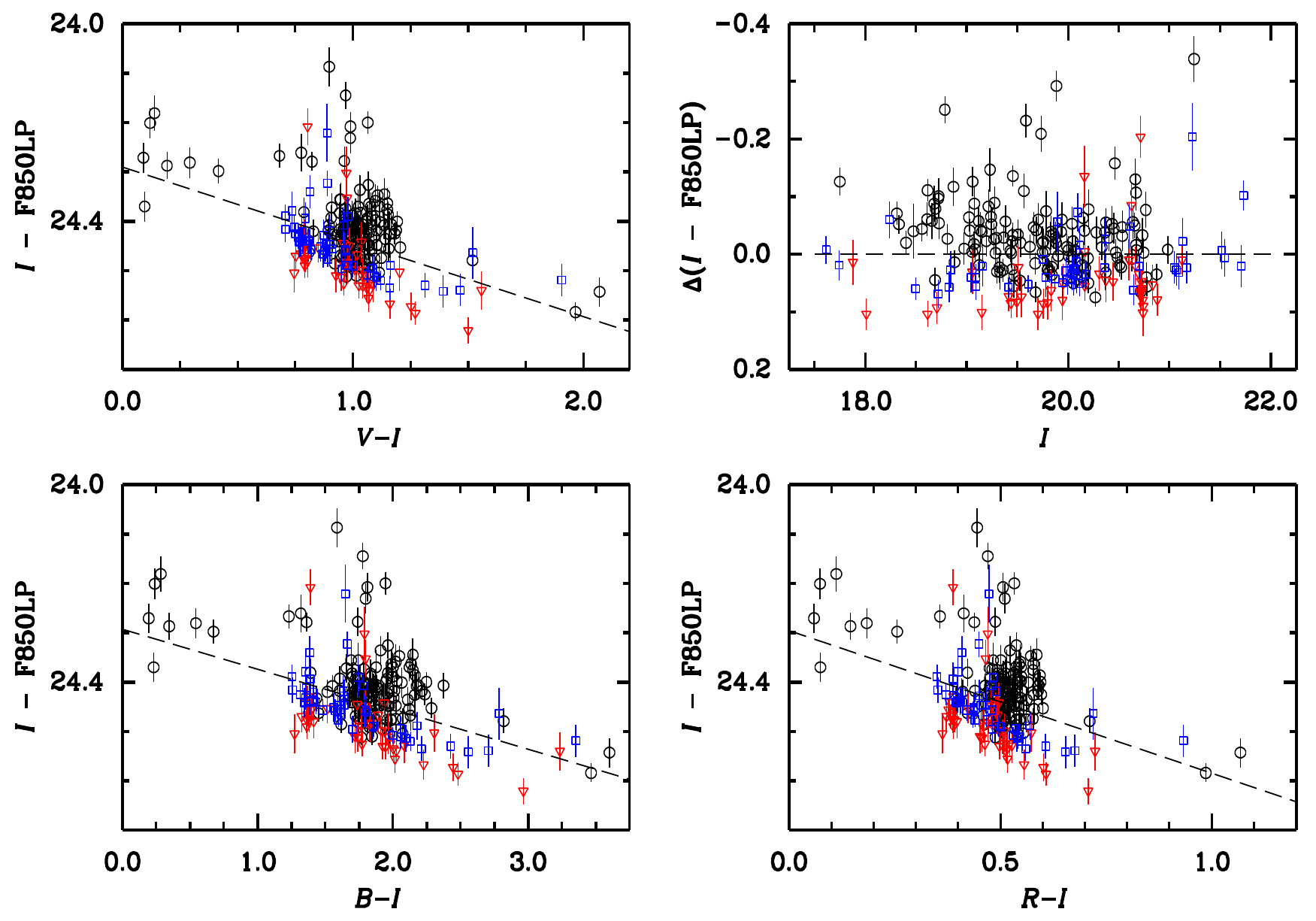}
\caption{Same as Figure~\ref{fig:f435w} for the filter F850LP. Residuals refer to the $I-F850LP$ vs. $V-I$ transformation.
See the electronic edition of the \textit{PASP} for a color version of this figure.
\label{fig:f850lp}}
\end{figure}

\begin{figure}
\epsscale{0.5}
\plotone{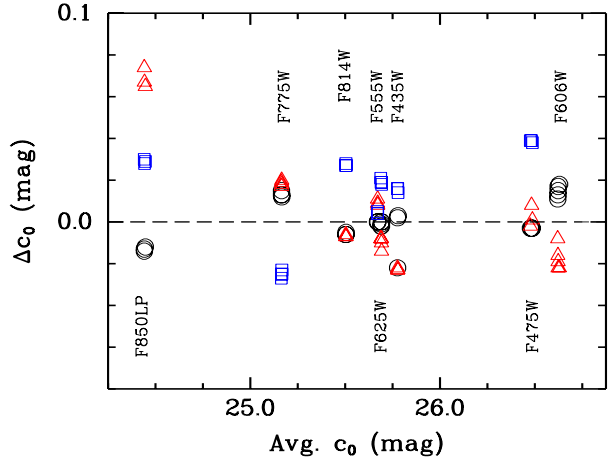}
\caption{Comparison of the derived zero-points of the photometric transformations (referenced to the weighted mean color of the stars: see text) as given in  Table~\ref{tab:PhotTrans}. 
Plotted are the differences in zero-point from the weighted mean as derived from the older data (NGC2419: \textit{black open circles}) and our more recent observations (PAL~4: \textit{red open triangles}; and PAL~14: \textit{blue open squares}) vs. the weighted mean photometric zero-point. Clusters of points at a common abscissa correspond to a particular filter, and except for F850LP show little deviation in the mean from zero (\textit{dashed line}). See the electronic edition of the \textit{PASP} for a color version of this figure.
\label{fig:PhZpt}}
\end{figure}

The principal goal of our program was to improve upon the seminal results of S05 using an approach that included observations of two additional globular clusters with somewhat different stellar populations (mean metallicity, age, and range of stellar colors), and to match to a much larger number of stars over a wider range of brightnesses. To assess the success of our strategy, we first consider the internal consistency of our results as presented in Figures~\ref{fig:f435w}--\ref{fig:f850lp}, where different symbols distinguish the contributions to the transformations from each target cluster. It is evident that the photometry is highly consistent with the derived linear transformations for all filters, with no need for higher-order terms. The transformations are also consistent for all the target clusters, except for the transformation for F850LP, which shows a small, constant offset between the data for different clusters. We examine this consistency in more detail in Figure~\ref{fig:PhZpt}, which compares the the zero-points of all the photometric transformations as derived from each cluster with their weighted-average. Here we make use of the zero-points referenced to the mean color for each transformation to remove the effect of uncertainties in the color term. It is clear that the dispersion of the zero-points for each filter is generally within 0.03~mag of the weighted mean, except for F850LP where the offset between NGC2419 and PAL4 is $\sim0.08$~mag; the origin of this discrepancy is not understood. 

We note the presence of several outliers in the transformations shown in Figs.~\ref{fig:f435w}--\ref{fig:f850lp}, the vast majority of which lie above the trend line by $\gtrsim0.15$~mag, which is to say that the instrumental magnitude is too faint relative to the ground-based Landolt magnitude. A visual inspection of the images at locations corresponding to $\sim10$ of the most deviant stars shows that most of the deviations in each passband are from the same stars, and the cause of the deviation is almost always either: 1) a close pair that was resolved with ACS, but would not have been resolved from the ground, or 2) part of the stellar profile was compromised by a detector artifact such as a bad column or a charge trap. Were these points to be excluded from the fits to the transformations, the effect would be of order 0.01~mag in $c_0$ (i.e., comparable to the formal error bars), and would be negligible in $c_1$. Since the effects are so small, we elected not to impose a posteriori rejection criteria on the data in order not to introduce more subtle systematics into the analysis. 
We also note that the transformations for three of the filters involve rather large color terms. The principal reason is that these filters are not especially well matched to the Landolt system. The F475W passband (SDSS $g$) spans the Landolt $B$ and $V$ passbands, and F606W spans Landolt $V$ and $R$. The F850LP passband (SDSS $z$) is strongly modulated by the decline in the CCD quantum efficiency in the red. Observations using these filters would not be a wise choice if the objective is to derive precision Landolt $BVRI$ photometry of stars, but we present the transformations for the benefit of archive users. Ultimately, though, the largest sources of error are likely systematic in nature. 

\begin{figure}
\epsscale{0.6}
\plotone{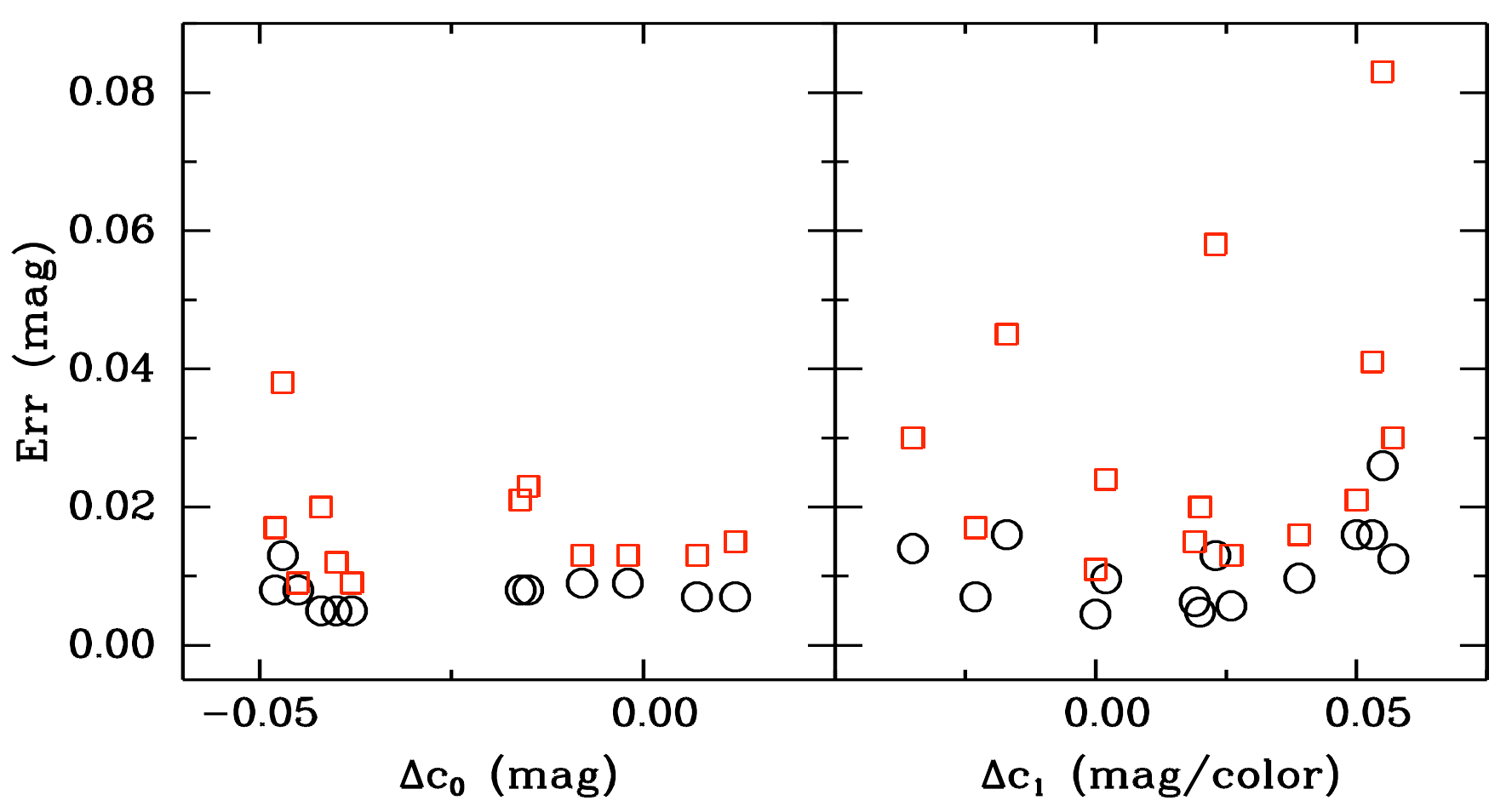}
\caption{Difference between the photometric transformation coefficients vs. the quoted uncertainties from this work (\textit{black circles}) and those of \citet{Sir05} (\textit{red squares}). \textit{Left:} difference between the zero-point terms (referenced to the weighted mean color of the stars: see text) vs. the quoted uncertainties; \textit{right:} difference between the color terms vs. the quoted uncertainties. 
See the electronic edition of the \textit{PASP} for a color version of this figure.
\label{fig:Error}}
\end{figure}

We now compare our transformations to those of S05 for filters in common, the coefficients for which are found in Table~\ref{tab:PhotTrans}. We find only small deviations in the $c_0$ terms of $\sim0.03$~mag for most transformations, which is illustrated in the left-hand panel of Figure~\ref{fig:Error}. The plot shows the difference between the combined $c_0$ terms derived here and those of S05 vs.\ the size of the quoted uncertainties. Note that these zero-points refer to those that apply at the mean color (designated $c^{\prime}_0$ in the tables) in order to minimize the impact of discrepancies in the $c_1$ terms.\footnote{We cannot determine what the uncertainties for the S05 $c^{\prime}_0$ terms would be at the mean color; we assume for this discussion that they are the same as for the $c_0$ terms.} The most discrepant points correspond to the transformations for the wide F606W filter, particularly for those involving Landolt $V$ and $R$. This is not surprising, and in part reflects an inherently higher uncertainty due to a strong dependency on the SED of the target star. The right-hand panel of Figure~\ref{fig:Error} shows the difference between the $c_1$ terms vs.\ their quoted uncertainties. It is clear that the color terms presented here have much reduced uncertainty, with a mean change in color that is for the most part consistent with the quoted errors of both studies. Again the largest changes in $c_1$ are for F606W, for the reasons noted above.

It is necessary to say something about the adopted Landolt magnitudes of the reference stars in the three globular clusters that were used in this work.  The estimated errors are described in detail by SDTW.  A detailed discussion is also given there concerning the discrepancies between the magnitudes of stars in NGC2419 as reported in SDTW vs. those {\it then} listed in the compilation on the Photometric Standard Fields website\footnote{The current website may be found at \url{http://www3.cadc-ccda.hia-iha.nrc-cnrc.gc.ca/community/STETSON/ standards/}.
} 
by \citet{Stetson00}. The above listings have since been modified by \citet{Stetson05}, where he found and corrected an error and revised the above listing.  This improves the discrepancy, but does not remove it entirely. There are still anomalies of $\sim 0.02$ mag between his listings  and our values for NGC2419, which are shown in detail in \citet{Stetson05}. Stetson, who had use of the image data that went into SDTW's published magnitudes, apparently got the same results as SDTW from this set of images, but these differ from the magnitudes he derived from other image data of NGC2419.  \citet{Stetson05} speculated that there is something wrong with the data used for SDTW, and specifically mentioned that an error in shutter timing of a few hundredths of a second could have generated the discrepancy. This is surprising, since SDTW had demonstrated  
that ``shutter timing and/or shading corrections are unnecessary at the 0.2\% level, even for exposures as short as 0.3~s," 
and included pointers to the commissioning report for the instrument (Mini-Mosaic on the WIYN 3.5-m telescope) to substantiate that claim. 
The remaining 0.02~mag discrepancy between the published mags of SDTW and those of Stetson's web listing for NGC2419 should be resolved with an independent data set with an independent analysis. Meanwhile the two authors of SDTW who are in common with those in this paper stand by the SDTW result,
though it is certainly possible for anyone to apply changes in the zero-point to be in accordance with \citet{Stetson05}. 

\subsection{Applicability of the Transformations}

The transformations we derived between the ACS/WFC system and that of Landolt are readily applicable to stars, and in particular stars with luminosity classes and metallicities represented in the calibrator sample considered here. In such cases the uncertainty in transforming between one system and the other can be well approximated from the uncertainties in the coefficients listed in Table~\ref{tab:PhotTrans}. The clusters observed for this study span a range of metallicity of a factor of $\sim4$, from $-1.5$~dex to $-2.1$~dex, relative to solar; the luminosity classes span main sequence, RGB, and AGB stars. S05 discussed at some length the additional uncertainties that apply when applying photometric transformations to stars that differ significantly in gravity or metallicity. S05 modeled the likely impact of these effects using synthetic photometry, spectral atlases, and model stellar atmospheres that spanned a large range in metallicity and spectral type (see their Figures 21 and 23). They provided alternative transformation coefficients, including a second-order color term, that best described their synthetic transformations. The veracity of these synthetic transformations depends upon the sampling of the input stellar atlases, the accuracy of the stellar atmosphere models, and the accuracy of the SYNPHOT synthetic photometry software, each of which can introduce uncertainties in the inferred transformations of a few to several percent. The differences between their empirical and synthetic transformations, which ranged from zero to $\sim0.05$~mag, depend in detail upon the filter used, the stellar gravity and metallicity, and the specific transformation considered. We believe that additional uncertainties of this order (a few to several percent) also apply to our empirical transformations for stellar targets that differ significantly in metallicity, gravity, and effective temperature from our sample of calibrator stars. 
Finally, we must caution readers that the transformations derived here are not likely to be accurate for non-stellar sources, including galaxies at high redshift. Our transformations have little meaningful applicability to sources with SEDs that contain strong emission, such as nebulae, symbiotic stars, active galaxies, and supernovae. 

The improvements in the transformations presented here from the ACS/WFC native system to that of Landolt are significant in a quantitative sense, particularly for the improved color terms. But the qualitative improvements are significant as well: comparing photometry from three different globular clusters gives a measure of systematic errors in the derivation, which on the whole appear to be comparable to the statistical uncertainties. Perhaps more significantly, since the target globular clusters were observed over an interval of about 3.5~yr, the agreement among the transformation coefficients indicates that the correction for the time-dependent CTE losses does not contribute significantly to the errors. The magnitude of the correction for faintest stars considered here is as large as $\sim0.3$~mag over the interval between epochs of the observations. Yet the agreement in the transformation coefficients is generally better than 0.03~mag. These results provide important reassurance that the accuracy of time-dependent CTE corrections \citep{chia09} is equal to the demands of high precision photometry. In the end we believe the comparison of our transformation relations among the three target clusters provides as good a sense of our external errors as can be obtained at this time, and that using the tabulated uncertainties in the coefficients provides a reasonable measure of the reliability of the photometric transformation for the variety of targets considered here. Finally, we presented results for two ACS/WFC filter bandpasses not reported on by S05, namely, F550M and F850LP. 

\acknowledgments

We would like to thank an anonymous referee, whose comments helped to improve this paper. Support for this work was provided by NASA through grants HST-GO-10621 and HST-GO-10622 from Space Telescope Science Institute, which is operated by the Association of Universities for Research in Astronomy, Inc., under NASA contract NAS~5-26555.


\clearpage


\begin{deluxetable}{llllrl}
\tabletypesize{\footnotesize}
\tablecolumns{6}
\tablewidth{0pt}
\tablecaption{ACS Observations \label{obslog}}

\tablehead{
\colhead{} & \colhead{} & \colhead{} & \colhead{} & \colhead{T$_{Exp}$} & \colhead{} \\*
\colhead{Object} & \colhead{UT date} & \colhead{Dataset} & \colhead{Filter} & \colhead{(s)} & \colhead{N$_{Exp}$\tablenotemark{a}}
}

\startdata
NGC2419\tablenotemark{b} & 2002 Sep 25 & \dataset[ADS/Sa.HST\#j8io01011]{j8io01011} & F435W  &  800 & 2 \\
        &             & \dataset[ADS/Sa.HST\#J8IO01021]{j8io01021} & F555W  &  720 & 2 \\
        &             & \dataset[ADS/Sa.HST\#j8io01031]{j8io01031} & F475W  &  680 & 2 \\
        &             & \dataset[ADS/Sa.HST\#j8io01041]{j8io01041} & F850LP & 1020 & 3 \\
        &             & \dataset[ADS/Sa.HST\#j8io01051]{j8io01051} & F775W  &  680 & 2 \\
        &             & \dataset[ADS/Sa.HST\#j8io01061]{j8io01061} & F625W  &  676 & 2 \\
        &             & \dataset[ADS/Sa.HST\#j8io01071]{j8io01071} & F814W  &  676 & 2 \\
        &             & \dataset[ADS/Sa.HST\#j8io01081]{j8io01081} & F606W  &  676 & 2 \\
 \\
PAL-4\tablenotemark{c} & 2006 Mar 06 & \dataset[ADS/Sa.HST\#j9bg01011]{j9bg01011} & F555W  &  250 & 2 \\
        &             & \dataset[ADS/Sa.HST\#j9bg01021]{j9bg01021} & F550M  &  500 & 2 \\
        &             & \dataset[ADS/Sa.HST\#j9bg01031]{j9bg01031} & F850LP &  380 & 2 \\
        &             & \dataset[ADS/Sa.HST\#j9bg01041]{j9bg01041} & F435W  &  430 & 2 \\
        &             & \dataset[ADS/Sa.HST\#j9bg01051]{j9bg01051} & F625W  &  180 & 2 \\
        &             & \dataset[ADS/Sa.HST\#j9bg01061]{j9bg01061} & F775W  &  210 & 2 \\
        &             & \dataset[ADS/Sa.HST\#j9bg01071]{j9bg01071} & F606W  &  120 & 2 \\
        &             & \dataset[ADS/Sa.HST\#j9bg01081]{j9bg01081} & F814W  &  160 & 2 \\
        &             & \dataset[ADS/Sa.HST\#j9bg01091]{j9bg01091} & F475W  &  230 & 2 \\
 \\
PAL-14\tablenotemark{c} & 2006 Apr 01 & \dataset[ADS/Sa.HST\#j9bg02011]{j9bg02011} & F555W  &  250 & 2 \\
        &             & \dataset[ADS/Sa.HST\#j9bg02021]{j9bg02021} & F550M  &  500 & 2 \\
        &             & \dataset[ADS/Sa.HST\#j9bg02031]{j9bg02031} & F850LP &  380 & 2 \\
        &             & \dataset[ADS/Sa.HST\#j9bg02041]{j9bg02041} & F435W  &  430 & 2 \\
        &             & \dataset[ADS/Sa.HST\#j9bg02051]{j9bg02051} & F625W  &  180 & 2 \\
        &             & \dataset[ADS/Sa.HST\#j9bg02061]{j9bg02061} & F775W  &  210 & 2 \\
        &             & \dataset[ADS/Sa.HST\#j9bg02071]{j9bg02071}\tablenotemark{d} & F606W  &  120 & 2\\
        &             & \dataset[ADS/Sa.HST\#j9bg02081]{j9bg02081} & F814W  &  160 & 2 \\
        &             & \dataset[ADS/Sa.HST\#j9bg02091]{j9bg02091} & F475W  &  230 & 2 \\
\enddata
\tablenotetext{a}{Exposure durations were CR-Split into sub-exposures of equal duration to facilitate cosmic-ray removal.} 
\tablenotetext{b}{Observed as part of ACS calibration program 9666.}
\tablenotetext{c}{Observed as part of GO program 10622.}
\tablenotetext{d}{{Loss} of guiding during observation: image unusable.}
\end{deluxetable}

\begin{deluxetable}{lcclccccc}
\tabletypesize{\tiny}
\tablecolumns{9}
\tablewidth{0pt}
\tablecaption{ACS/WFC to $BVRI$ Photometric Transformations \label{tab:PhotTrans}}
\tablehead{
\colhead{$SMAG$} & \colhead{$TMAG$} & \colhead{$TCOL$} & \colhead{Coefficient} & \colhead{NGC2419} & \colhead{PAL4} & \colhead{PAL14} & \colhead{Combined} & \colhead{Sirianni\tablenotemark{a}}}
\startdata
	& 		&		& No. Stars 		& 211				& 39					& 56					& 311				& \ldots \\ 
F435W & $B$	& $B-V$ 	& Mean Color		& 0.904				& 0.871				& 0.784				& 0.873				& \ldots \\*
	&		&		& $c_{0}$			& $25.867\pm0.019$	& $25.821\pm0.020$	& $25.840\pm0.015$	& $25.853\pm0.016$	& $25.842\pm0.023$ \\*
	& 		& 		& $c_{1}$			& $-0.100\pm0.010$		& $-0.077\pm0.011$		& $-0.055\pm0.009$		& $-0.087\pm0.009$		& $-0.089\pm0.024$ \\*
	& 		& & $c^{\prime}_0$($B-V=0.85$) & $25.782\pm0.009$ 	& $25.756\pm0.010$ 	& $25.793\pm0.007$ 	& $25.779\pm0.008$	& 25.766 \\ 
\\
	& 		& $B-R$ 	& Mean Color		& 1.432				&  1.390				&  1.262				&  1.389				& \ldots \\*
	&		&		& $c_{0} $			& $25.869\pm0.019$	& $25.828 \pm0.020$	& $25.839\pm0.015$	&  $25.853\pm0.016$	& \ldots \\*
	&		&		& $c_{1} $			& $-0.065\pm0.006$		& $-0.053\pm0.007$		& $-0.034\pm0.006$		&  $-0.057\pm0.006$	& \ldots \\*
	&		& & $c^{\prime}_0$($B-R=1.40$) & $25.778\pm0.009$ 	& $25.754\pm0.010$ 	& $25.792\pm0.008$	& $25.776\pm0.008$ 	& \ldots \\ 
\\
	& 		& $B-I$ 	& Mean Color		& 1.966				& 1.879				& 1.735				& 1.906				& \ldots \\*
	&		&		& $c_{0} $			& $25.872\pm0.019$	& $25.834\pm0.020$	& $25.840\pm0.015$	& $25.858\pm0.016$	& $25.847\pm0.021$ \\*
	&		&		& $c_{1} $			& $-0.049\pm0.005$		& $-0.042\pm0.005$		& $-0.025\pm0.004$		& $-0.043\pm0.004$		& $-0.045\pm0.011$ \\*
	&		& & $c^{\prime}_0$($B-I=1.90$) & $25.779\pm0.010$	& $25.754\pm0.010$	& $25.793\pm0.008$	& $25.777\pm0.008$	& 25.762\\ 
\\
	&		&		& No. Stars 		& 205				& 39					& 56					&  312				& \ldots \\*
F475W & $B$	& $B-V$ 	& Mean Color		& 0.887				& 0.863				& 0.793				& 0.868				& \ldots \\*
	&		&		& $c_{0}$			& $26.182\pm0.020$	& $26.092\pm0.008$	& $26.163\pm0.012$	& $26.177\pm0.020$	& \ldots \\*
	&		&		& $c_{1}$			& $0.343\pm0.011$		& $0.451\pm0.004$		& $0.416\pm0.007$		& $0.353\pm0.011$		& \ldots \\*
	&		& & $c^{\prime}_0$($B-V=0.85$) & $26.474\pm0.010$	& $26.475\pm0.004$	& $26.516\pm0.006$	& $26.477\pm0.010$	& \ldots \\
\\
	& 		& $B-R$ 	& Mean Color		& 1.409				&  1.378				&  1.273				& 1.382	& \ldots \\*
	&		&		& $c_{0}$			& $26.188\pm0.021$	& $26.046\pm0.008$	& $26.163\pm0.014$	& $26.174\pm0.021$	& \ldots \\*
	&		&		& $c_{1}$			& $0.212\pm0.007$		& $0.316\pm0.003$		& $0.259\pm0.005$		& $0.223\pm0.007$		& \ldots \\*
	&		& & $c^{\prime}_0$($B-R=1.40$) & $26.484\pm0.010$	& $26.488\pm0.004$	& $26.525\pm0.007$	& $26.487\pm0.011$	& \ldots \\ 
\\
	& 		& $B-I$ 	& Mean Color		& 1.938				&  1.863				&  1.752				& 1.897				& \ldots \\*
	&		&		& $c_{0}$			& $26.177\pm0.022$	& $26.007\pm0.008$	& $26.142\pm0.014$	& $26.161\pm0.022$	& \ldots \\*
	&		&		& $c_{0}$			& $0.159\pm0.005$		& $0.254\pm0.002$		& $0.200\pm0.004$		& $0.170\pm0.006$		& \ldots \\*
	&		& & $c^{\prime}_0$($B-I=1.90$) & $26.480\pm0.011$	& $26.491\pm0.004$	& $26.522\pm0.007$	& $26.483\pm0.011$	& \ldots \\ 
\\
	&		&		& No. Stars 		& \ldots				& 42					& 52					& 99					& \ldots \\*
F550M & $V$	& $B-V$	& Mean Color		& \ldots				& 0.840				& 0.786				& 0.808				& \ldots \\* 
	&		&		& $c_{0}$			& \ldots				& $24.875\pm0.018$	& $24.847\pm0.012$	& $24.852\pm0.021$	& \ldots \\* 
	&		&		& $c_{1}$			& \ldots				& $0.077\pm0.010$		& $0.145\pm0.007$		& $0.117\pm0.013$		& \ldots \\* 
	&		& & $c^{\prime}_0$($B-V=0.85$) & \ldots				& $24.941\pm0.009$	& $24.970\pm0.006$	& $24.951\pm0.011$	& \ldots \\ 
\\
	& 		& $V-R$ 	& Mean Color		& \ldots				& 0.504				& 0.474				& 0.486				& \ldots \\* 
	&		&		& $c_{0}$			& \ldots				& $24.844\pm0.018$	& $24.835\pm0.011$	& $24.829\pm0.020$	& \ldots \\* 
	&		&		& $c_{1}$			& \ldots				& $0.190\pm0.017$		& $0.265\pm0.011$		& $0.240\pm0.021$		& \ldots \\* 
	&		& & $c^{\prime}_0$($V-R=0.50$) & \ldots				& $24.939\pm0.009$	& $24.967\pm0.006$	& $24.949\pm0.010$	& \ldots \\ 
\\
	& 		& $V-I$ 	& Mean Color		& \ldots				& 0.982				& 0.947				& 0.961				& \ldots  \\* 
	&		&		& $c_{0}$			& \ldots				& $24.836\pm0.018$	& $24.819\pm0.011$	& $24.814\pm0.020$	& \ldots \\* 
	&		&		& $c_{1}$			& \ldots				& $0.106\pm0.009$		& $0.150\pm0.006$		& $0.138\pm0.010$		& \ldots \\* 
	&		& & $c^{\prime}_0$($V-I)=1.00$) & \ldots				& $24.942\pm0.009$	& $24.968\pm0.005$	& $24.951\pm0.010$	& \ldots  \\ 
\\
	&		&		& No. Stars 		& 187				&  41					&  55					& 291				& \ldots \\*
F555W & $V$	& $B-V$ 	& Mean Color		& 0.857				&  0.876				&  0.790				&  0.849				& \ldots \\* 
	&		&		& $c_{0}$			& $25.740\pm0.011$	& $25.733\pm0.006$	& $25.740\pm0.013$	&$25.740\pm0.010$		& $25.701\pm0.012$ \\* 
	&		&		& $c_{1}$			& $-0.080\pm0.007$		& $-0.062\pm0.004$		& $-0.076\pm0.008$		& $-0.081\pm0.006$		& $-0.056\pm0.013$ \\* 
	&		& & $c^{\prime}_0$($B-V=0.85$) & $25.671\pm0.006$	& $25.680\pm0.003$	& $25.675\pm0.007$	& $25.671\pm0.005$	& 25.653\\ 
\\
	& 		& $V-R$ 	& Mean Color		& 0.506				&  0.520				&  0.477				& 0.503				& \ldots \\* 
	&		&		& $c_{0}$			& $25.736\pm0.012$	& $25.755\pm0.006$	& $25.740\pm0.014$	& $25.740\pm0.010$	& $25.703\pm0.009$ \\* 
	&		&		& $c_{1}$			& $-0.129\pm0.011$		& $-0.146\pm0.006$		& $-0.127\pm0.014$		& $-0.137\pm0.010$		& $-0.100\pm0.016$ \\* 
	&		& & $c^{\prime}_0$($V-R=0.50$) & $25.672\pm0.006$ 	& $25.682\pm0.003$	& $25.676\pm0.007$	& $25.671\pm0.005$	& 25.653\\ 
\\
	& 		& $V-I$ 	& Mean Color		& 1.023				& 1.011				& 0.951				& 1.011				& \ldots \\* 
	&		&		& $c_{0}$			& $25.739\pm0.012$	& $25.7601\pm0.006$	& $25.746\pm0.013$	& $25.745\pm0.010$	& $25.704\pm0.020$ \\* 
	&		&		& $c_{1}$			& $-0.067\pm0.006$		& $-0.080\pm0.003$		& $-0.070\pm0.007$		& $-0.073\pm0.005$		& $-0.054\pm0.020$ \\* 
	&		& & $c^{\prime}_0$($V-I=1.00$) & $25.672\pm0.006$	& $25.680\pm0.003$	& $25.676\pm0.007$	& $25.672\pm0.005$	& 25.650\\ 
\\
	&		&		& No. Stars 		& 123				& 39					& \ldots				& 170				& \ldots \\*
F606W & $V$	& $B-V$	& Mean Color		& 0.788				& 0.854				& \ldots				& 0.803				& \ldots \\* 
	&		&		& $c_{0}$			& $26.427\pm0.012$	& $26.471\pm0.053$	& \ldots				& $26.444\pm0.026$	& $26.399\pm0.038$ \\* 
	&		&		& $c_{1}$			& $0.250\pm0.007$		& $0.156\pm0.031$		& \ldots				& $0.211\pm0.016$		& $0.193\pm0.045$ \\*
	&		& & $c^{\prime}_0$($B-V=0.85$) & $26.640\pm0.006$	& $26.604\pm0.027$ 	& \ldots				& $26.623\pm0.013$	& 26.563\\ 
\\
	& 		& $B-R$ 	& Mean Color		& 1.254				& 1.364				& \ldots				& 1.280				& \ldots \\* 
	&		&		& $c_{0}$			& $26.424\pm0.011$	& $26.456\pm0.054$	& \ldots				& $26.435\pm0.025$	& \ldots \\* 
	&		&		& $c_{1}$			& $0.160\pm0.004$		& $0.109\pm0.020$		& \ldots				& $0.139\pm0.010$		& \ldots \\* 
	&		& & $c^{\prime}_0$($B-R=1.40$) & $26.648\pm0.006$	& $26.608\pm0.027$ 	& \ldots				& $26.630\pm0.013$	& \ldots \\ 
\\
	& 		& $V-R$ 	& Mean Color		& 0.466				& 0.510				& \ldots				& 0.477				& \ldots \\* 
	&		&		& $c_{0}$			& $26.427\pm0.011$	& $26.421\pm0.056$	& \ldots				& $26.428\pm0.025$	& $26.341\pm0.041$ \\* 
	&		&		& $c_{1}$			& $0.423\pm0.011$		& $0.360\pm0.054$		& \ldots				& $0.389\pm0.026$		& $0.444\pm0.083$ \\* 
	&		& & $c^{\prime}_0$($V-R=0.50$) & $26.638\pm0.005$ 	& $26.601\pm0.028$	& \ldots				& $26.623\pm0.013$	& 26.563 \\ 
\\
	& 		& $V-I$ 	& Mean Color		& 0.950				&  0.992				& \ldots				& 0.960				& \ldots \\* 
	&		&		& $c_{0}$			& $26.414\pm0.011$	& $26.402\pm0.055$	& \ldots				& $26.410\pm0.025$	& $26.325\pm0.057$ \\* 
	&		&		& $c_{1}$			& $0.222\pm0.006$		& $0.204\pm0.027$		& \ldots				& $0.212\pm0.013$		& $0.236\pm0.058$ \\* 
	&		& & $c^{\prime}_0$($V-I=1.00$) & $26.635\pm0.005$	& $26.606\pm0.027$	& \ldots				& $26.622\pm0.013$	& 26.561\\ 
\\
	& 		& $R-I$ 	& Mean Color		& 0.484				& 0.482				& \ldots				& 0.484				& \ldots \\* 
	&		&		& $c_{0}$			& $26.404\pm0.011$	& $26.382\pm0.054$	&\ldots				& $26.397\pm0.025$	& \ldots \\* 
	&		&		& $c_{1}$			& $0.455\pm0.011$		& $0.463\pm0.056$		& \ldots				& $0.447\pm0.026$		& \ldots \\* 
	&		& & $c^{\prime}_0$($R-I=0.50$) & $26.632\pm0.006$	& $26.613\pm0.027$	& \ldots				& $26.621\pm0.013$	& \ldots \\
\\
	&		&		& No. Stars 		& 141				& 41					& 51					& 247				& \ldots \\*
F625W & $R$	& $B-V$ 	& Mean Color		& 0.805				& 0.843				& 0.787				& 0.806				& \ldots \\*
	&		&		& $c_{0}$			& $25.764\pm0.015$	& $25.735\pm0.011$	& $25.750\pm0.011$	& $25.759\pm0.016$	& \ldots \\*
	&		&		& $c_{1}$			& $-0.088\pm0.009$		& $-0.062\pm0.007$		& $-0.049\pm0.007$		& $-0.080\pm0.010$		& \ldots \\*
	&		& & $c^{\prime}_0$($B-V=0.85$) & $25.689\pm0.008$	& $25.682\pm0.006$	& $25.709\pm0.006$	&$25.690\pm0.008$		& \ldots \\ 
\\
	& 		& $B-R$ 	& Mean Color		& 1.282				& 1.347				& 1.264				& 1.286				& \ldots \\*
	&		&		& $c_{0}$			& $25.766\pm0.015$	& $25.743\pm0.011$	& $25.753\pm0.011$	& $25.762\pm0.016$	& \ldots \\*
	&		&		& $c_{1}$			& $-0.057\pm0.006$		& $-0.045\pm0.004$		& $-0.033\pm0.004$		& $-0.053\pm0.006$		& \ldots \\*
	&		& & $c^{\prime}_0$($B-R=1.40$) & $25.686\pm0.008$	& $25.680\pm0.006$	& $25.709\pm0.005$	& $25.688\pm0.008$	& \ldots \\ 
\\
	& 		& $V-R$ 	& Mean Color		& 0.477				& 0.504				& 0.477				& 0.481				& \ldots \\*
	&		&		& $c_{0}$			& $25.766\pm0.015$	& $25.762\pm0.011$	& $25.757\pm0.010$	& $25.764\pm0.016$	& $25.720\pm0.009$ \\*
	&		&		& $c_{1}$			& $-0.154\pm0.015$		& $-0.158\pm0.011$		& $-0.094\pm0.011$		& $-0.146\pm0.016$		& $-0.098\pm0.021$ \\*
	&		& & $c^{\prime}_0$($V-R=0.50$) & $25.689\pm0.007$	& $25.683\pm0.006$	& $25.710\pm0.005$	& $25.691\pm0.008$	& 25.671\\ 
\\
	& 		& $V-I$ 	& Mean Color		& 0.970				& 0.982				& 0.953				& 0.968				& \ldots \\*
	&		&		& $c_{0}$			& $25.771\pm0.015$	& $25.766\pm0.011$	& $25.758\pm0.011$	& $25.769\pm0.016$	& \ldots \\*
	&		&		& $c_{1}$			& $-0.080\pm0.008$		& $-0.085\pm0.006$		& $-0.049\pm0.006$		& $-0.078\pm0.008$		& \ldots \\*
	&		& & $c^{\prime}_0$($V-I=1.00$) & $25.691\pm0.007$	& $25.681\pm0.006$	& $25.710\pm0.005$	& $25.691\pm0.008$	& \ldots \\ 
\\
	& 		& $R-I$ 	& Mean Color		& 0.494				& 0.477				& 0.476				& 0.487				& \ldots \\*
	&		&		& $c_{0}$			& $25.774\pm0.015$	& $25.769\pm0.011$	& $25.759\pm0.011$	& $25.772\pm0.016$	& $25.726\pm0.017$ \\*
	&		&		& $c_{1}$			& $-0.165\pm0.015$		& $-0.182\pm0.012$		& $-0.099\pm0.011$		& $-0.161\pm0.016$		& $-0.110\pm0.041$ \\*
	&		& & $c^{\prime}_0$($R-I=0.50$) & $25.692\pm0.008$	& $25.678\pm0.006$	& $25.710\pm0.006$	& $25.692\pm0.008$	& 25.671\\
\\
	&		&		& No. Stars 		& 100				& 37					& 43					& 190				& \ldots \\ 
F775W & $I$	& $B-I$ 	& Mean Color		&  1.776				& 1.834				& 1.716				& 1.769				& \ldots \\* 
	&		&		& $c_{0}$			& $25.247\pm0.010$	& $25.244\pm0.009$	& $25.273\pm0.029$	&$ 25.230\pm0.017$	& \ldots \\* 
	&		&		& $c_{1}$			& $-0.037\pm0.003$		& $-0.032\pm0.002$		& $-0.072\pm0.008$		& $-0.035\pm0.005$		& \ldots \\* 
	&		& & $c^{\prime}_0$($B-I=1.90$) & $25.177\pm0.005$	& $25.183\pm0.004$	& $25.137\pm0.015$	& $25.164\pm0.009$	& \ldots \\*
\\
	& 		& $V-R$ 	& Mean Color		&  0.477				&  0.508				&  0.469				&  0.480				& \ldots \\* 
	&		&		& $c_{0}$			& $25.244\pm0.010$	& $25.253\pm0.009$	& $25.263\pm0.030$	& $25.230\pm0.017$	& \ldots \\* 
	&		&		& $c_{1}$			& $-0.132\pm0.010$		& $-0.134\pm0.009$		& $-0.241\pm0.031$		& $-0.129\pm0.018$		& \ldots \\* 
	&		& & $c^{\prime}_0$($V-R=0.50$) & $25.178\pm0.005$	& $25.186\pm0.004$	& $25.143\pm0.015$	& $25.166\pm0.009$	& \ldots \\ 
\\
	& 		& $V-I$ 	& Mean Color		&  0.968				&  0.987				&  0.939				&  0.963				& \ldots \\* 
	&		&		& $c_{0}$			& $25.250\pm0.010$	& $25.257\pm0.009$	& $25.286\pm0.029$	& $25.240\pm0.017$	&$25.240\pm0.013$ \\* 
	&		&		& $c_{1}$			& $-0.071\pm0.005$		& $-0.073\pm0.009$		& $-0.145\pm0.015$		& $-0.074\pm0.009$		& $-0.054\pm0.015$ \\* 
	&		& & $c^{\prime}_0$($V-I=1.00$) & $25.179\pm0.005$	& $25.184\pm0.004$	& $25.141\pm0.015$	& $25.166\pm0.009$	& 25.186\\ 
\\
	& 		& $R-I$	& Mean Color		&  0.491				&  0.480				&  0.470				&  0.483				& \ldots \\*
	&		&		& $c_{0}$			& $25.256\pm0.010$	& $25.261\pm0.009$	& $25.312\pm0.028$	& $25.250\pm0.017$	& $25.242\pm0.013$ \\* 
	&		&		& $c_{1}$			& $-0.151\pm0.010$		& $-0.157\pm0.009$		& $-0.344\pm0.029$		& $-0.169\pm0.017$		& $-0.106\pm0.030$ \\* 
	&		& & $c^{\prime}_0$($R-I=0.50$) & $25.180\pm0.005$	& $25.182\pm0.004$	& $25.140\pm0.014$	& $25.165\pm0.009$	& 25.189 \\
\\
	&		&		& No. Stars 		& 135				& 35					& 47					& 223				& \ldots \\*
F814W & $I$	& $B-I$ 	& Mean Color		& 1.745				&  1.762				&  1.690				&  1.735				& \ldots \\* 
	&		&		& $c_{0}$			& $25.470\pm0.020$	& $25.450\pm0.009$	& $25.510\pm0.007$	& $25.479\pm0.014$	& \ldots \\* 
	&		&		& $c_{1}$			& $0.016\pm0.006$		& $0.025\pm0.002$		& $0.012\pm0.002$		& $0.014\pm0.004$		& \ldots \\* 
	&		& & $c^{\prime}_0$($B-I=1.90$) & $25.500\pm0.010$	& $25.498\pm0.004$	& $25.532\pm0.004$	& $25.505\pm0.007$	& \ldots \\ 
\\
	& 		& $V-I$ 	& Mean Color		&  0.956				&  0.957				&  0.925				& 0.949				& \ldots \\* 
	&		&		& $c_{0}$			& $25.472\pm0.0120$	& $25.442\pm0.009$	& $25.510\pm0.007$	& $25.480\pm0.014$	& $25.495\pm0.015$ \\* 
	&		&		& $c_{1}$			& $0.026\pm0.010$		& $0.055\pm0.005$		& $0.022\pm0.004$		& $0.024\pm0.007$		& $-0.002\pm0.017$ \\* 
	&		& & $c^{\prime}_0$($V-I=1.00$) & $25.498\pm0.010$	& $25.497\pm0.004$	& $25.532\pm0.004$	& $25.504\pm0.007$	& 25.494\\ 
\\
	& 		& $R-I$ 	& Mean Color		&  0.486				&  0.466				&  0.465				&  0.478				& \ldots \\* 
	&		&		& $c_{0}$			& $25.475\pm0.020$	& $25.442\pm0.009$	& $25.517\pm0.007$	& $25.482\pm0.014$	& $25.492\pm0.013$ \\* 
	&		&		& $c_{1}$			& $0.046\pm0.020$		& $0.113\pm0.009$		& $0.028\pm0.008$		& $0.043\pm0.014$		& $0.002\pm0.030$ \\* 
	&		& & $c^{\prime}_0$($R-I=0.50$) & $25.498\pm0.010$	& $25.498\pm0.004$	& $25.531\pm0.004$	& $25.504\pm0.007$	& 25.493\\
\\
	&		&		& No. Stars		& 132				& 38					& 45					& 224				& \ldots \\*
F850LP & $I$	& $B-I$ 	& Mean Color		& 1.834				& 1.818				& 1.731				& 1.812				& \ldots \\* 
	&		&		& $c_{0}$			& $24.280\pm0.017$	& $24.351\pm0.009$	& $24.314\pm0.012$	& $24.293\pm0.021$	& \ldots \\* 
	&		&		& $c_{1}$			& $0.082\pm0.005$		& $0.085\pm0.003$		& $0.086\pm0.004$		& $0.081\pm0.006$		& \ldots \\* 
	&		& & $c^{\prime}_0$($B-I=1.90$) & $24.435\pm0.009$	& $24.512\pm0.005$	& $24.476\pm0.006$	& $24.447\pm0.011$	& \ldots \\ 
\\
	& 		& $V-I$ 	& Mean Color		&  1.003				&  0.980				&  0.947				&  0.990				& \ldots \\* 
	&		&		& $c_{0}$			& $24.276\pm0.017$	& $24.314\pm0.009$	& $24.312\pm0.014$	& $24.290\pm0.021$	& \ldots \\* 
	&		&		& $c_{1}$			& $0.154\pm0.009$		& $0.195\pm0.005$		& $0.158\pm0.007$		& $0.151\pm0.011$		& \ldots \\* 
	&		& & $c^{\prime}_0$($V-I=1.00$) & $24.429\pm0.009$	& $24.509\pm0.005$	& $24.470\pm0.007$	& $24.442\pm0.011$	& \ldots \\ 
\\
	& 		& $R-I$ 	& Mean Color		&  0.510				&  0.476				&  0.472				&   0.498				& \ldots \\* 
	&		&		& $c_{0}$			& $24.271\pm0.018$	& $24.305\pm0.009$	& $24.300\pm0.014$	& $24.296\pm0.022$	& \ldots \\* 
	&		&		& $c_{1}$			& $0.312\pm0.017$		& $0.421\pm0.010$		& $0.343\pm0.015$		& $0.289\pm0.022$		& \ldots \\* 
	&		& & $c^{\prime}_0$($R-I=0.50$) & $24.427\pm0.009$	& $24.515\pm0.005$	& $24.471\pm0.007$	& $24.441\pm0.011$	& \ldots \\
\enddata
\tablenotetext{a}{Transformations from \citet{Sir05}.} 
\end{deluxetable}



\clearpage

\end{document}